\documentclass[aps,amsfonts,twocolumn,nofootinbib,floatfix]{revtex4}
\usepackage{amsmath}
\usepackage{epsfig}
\usepackage{latexsym}
\usepackage{color}

\def\be{\begin{equation}}
 \def\ee{\end{equation}}
 \def\bea{\begin{eqnarray}}
 \def\eea{\end{eqnarray}}

\def\2{\frac{1}{2}}
\def\4{\frac{1}{4}}

\def\gen{\mathrm{g}}

\catcode`\@=11

\catcode`\@=12

\begin{document}

\title{Shear Viscosity and Chern-Simons Diffusion Rate from Hyperbolic Horizons}

\author{George Koutsoumbas$^{1}$,
Eleftherios Papantonopoulos$^{1}$,  George Siopsis$^{2}$}
\affiliation{\vspace*{0.2cm}$^1$Department of Physics, National
Technical University of Athens, GR~157~73~Athens, Greece}
\affiliation{ $^2$Department of Physics and Astronomy, The
University of Tennessee, Knoxville, TN 37996 - 1200, USA}
\date{\today}

\begin{abstract}

We calculate the shear viscosity and anomalous baryon number
violation rate in quantum field theories at finite temperature
having gravity duals with hyperbolic horizons. We find the
explicit dependence of these quantities on the temperature. We
show that the ratio of shear viscosity to entropy density is below
$1/(4\pi)$ at all temperatures and can be made arbitrarily small
in the low temperature limit for hyperbolic surfaces of
sufficiently high genus so that the hydrodynamic limit remains
valid.

\end{abstract}

\maketitle

In certain finite-temperature quantum field theories having
gravity duals with black brane solutions in higher spacetime
dimensions, the hydrodynamic behaviour of the thermal field theory
is identified with the hydrodynamic behaviour of the dual gravity
theory~\cite{Hydrodynamics}. It was shown~\cite{Kovtun:2004de}
that for these field theories, the ratio of the shear viscosity to
the volume density of entropy has a universal value
$\eta/s=1/(4\pi)$ and it was further conjectured that this is the
lowest bound on the ratio $\eta/s$ for a large class of thermal
quantum field theories.

This conjecture was tested against a wide range of thermal field
theories having gravity duals: in gauge theories with chemical
potentials studying their R-charged black hole
duals~\cite{chemical}, in field theories with stringy
corrections~\cite{corrections} and also in field theories with
gravity duals of Einstein-Born-Infeld gravity~\cite{Cai:2008in}.
In all these theories  it was found that the lower bound is
satisfied. However, in conformal field theories dual to Einstein
gravity with curvature square corrections it was found that the
bound is violated~\cite{GB_correction} but the physical
implication of the violation of the bound is still not clear.

In the gravity sector of this gravity/gauge duality, maximally
symmetric spaces naturally arise as the near-horizon region of
black brane geometries~\cite{Horowitz:1991cd}. Spherically
symmetric spaces have been extensively investigated. Also
hyperbolic geometries involving $n$-dimensional hyperboloids
$\mathbb{H}^n$ or $\mathbb{H}^n/\Gamma $ cosets, where $\Gamma$ is
a discrete subgroup of the isometry group of $\mathbb{H}^n$, arise
naturally in supergravity as a result of string
compactifications~\cite{Kehagias:2000dga}. However, the presence
of the discrete group $\Gamma $ introduces another scale which
breaks all supersymmetries. $\mathcal{N}=0$ conformal field
theories can be constructed having gravity duals with constant
negative curvature~\cite{Emparan:1999gf,Kehagias:2000dga}.

The hydrodynamic properties of the boundary conformal field theory
can be inferred from the lowest frequency quasinormal modes of the
gravity sector~\cite{Son:2002sd}.  The lowest-lying gravitational
quasinormal modes for a Schwarzschild-AdS solution were
numerically calculated in four and five dimensions and were shown
to be in agreement with hydrodynamic perturbations of the gauge
theory plasma on the AdS boundary~\cite{Friess:2006kw}.  For
AdS$_5$ this was understood as a finite ``conformal soliton flow"
after the spherical AdS$_5$ boundary obtained in global
coordinates was conformally mapped to the physically relevant flat
Minkowski spacetime. This study was extended to  black holes with
a hyperbolic horizon. It was shown in~\cite{Alsup:2008fr} that the
quasinormal modes obtained agreed with the frequencies resulting
from considering perturbations of the gauge theory fluid on the
boundary.

Recently, interesting features have shown up in the study of
topological black holes (TBH). The spectrum of the quasinormal
modes of TBH~\cite{topological} has been studied
extensively~\cite{TBH_perturbations}. For large black holes this
spectrum is similar to the Schwarzschild-AdS spectrum. For small
black holes however the quasinormal modes spectrum is quite
different. It was found~\cite{Koutsoumbas:2006xj} that there is a
critical temperature, below which there is a phase transition of
the TBH to AdS space. This has been attributed entirely to the
properties of the hyperbolic geometry.

In this work we will show that the hyperbolic geometry allows us
to calculate hydrodynamic transport coefficients like shear
viscocity and the Chern-Simons diffusion rate of the boundary
thermal field theory at any temperature under certain conditions. This should
be constrasted with
the case of a spherical black hole where low temperature is invariably associated with small horizon area and therefore
the hydrodynamic approximation breaks down. In the hyperbolic case,
the area of the horizon can be large even at low temperatures provided the hyperbolic surface is of high genus.

Topological black holes are solutions of the Einstein equations
for vacuum AdS space. Consider the action
\begin{equation}
I=\frac{1}{16\pi G}\int d^{d}x \sqrt{-g}
\biggl[R+\frac{(d-1)(d-2)}{l^{2}}\biggr]~,
\end{equation}
 where $G$ is the Newton's constant,
$R$ is the Ricci scalar and $l$ is the AdS radius. The presence of
a negative cosmological constant $(\Lambda =-
\frac{(d-1)(d-2)}{2l^{2}})$ allows the existence of black holes
with topology $\mathbb{R}^{2}\times\Sigma$, where $\Sigma$ is a
$(d-2)$-dimensional manifold of constant negative curvature. These
black holes are known as topological black holes
(TBHs)~\cite{topological}. The simplest solution of this kind in
four dimensions reads \bea
ds^{2}&=&-f(r)dt^{2}+\frac{1}{f(r)}dr^{2}+r^{2}d\sigma
^{2}\nonumber \\
 f(r)&=&r^{2}-1-2G\mu /r\label{linel}~,
\eea where  we have set the AdS radius $l=1$, $\mu$ is a constant
which is proportional to the mass  and $d\sigma^{2}$ is the line
element of the two-dimensional manifold $\Sigma$, which is locally
isomorphic to the hyperbolic manifold $\mathbb{H}^{2}$ and of the form
\begin{equation}
\Sigma=\mathbb{H}^{2}/\Gamma \quad \textrm{,\quad  $\Gamma\subset
O(2,1)$}~,
\end{equation}
where $\Gamma$ is a freely acting discrete subgroup (i.e. without
fixed points) of isometries. The line element $d\sigma^{2}$ of
$\Sigma$ is
\begin{equation}
d\sigma^{2}=d\theta^{2}+\sinh^{2}\theta d\varphi{^2}~,
\end{equation}
with $\theta\ge0$ and $0\le\phi<2\pi$ being the coordinates of the
hyperbolic space $\mathbb{H}^{2}$ or pseudosphere, which is a non-compact
two-dimensional space of constant negative curvature. This space
becomes a compact space of constant negative curvature with genus
$\gen\ge2$ by identifying, according to the connection rules of
the discrete subgroup $\Gamma$, the opposite edges of a
$4\gen$-sided polygon whose sides are geodesics and is centered at
the origin $\theta=\varphi=0$ of the
pseudosphere~\cite{topological,HyperB}. An octagon is the simplest
such polygon, yielding a compact surface of genus $\gen=2$ under
these identifications. Thus, the two-dimensional manifold $\Sigma$
is a compact Riemann 2-surface of genus $\mathrm{g}\geq2$.  The
configuration (\ref{linel}) is an asymptotically locally AdS
spacetime.

This construction can be generalized to higher dimensions and our
aim in this work is to elucidate the effect of hyperbolic horizons
on the gauge theory on the AdS boundary. In five spacetime
dimensions the metric takes the form \be\label{metric} ds^2=- f(r)
dt^2+\frac{dr^2}{f(r)}+r^2d\Sigma^2_3 \ \ , \ \ \ \ f(r) = r^2 -1
- \frac{2\mu}{r^2}~, \ee where $\Sigma_3 = \mathbb{H}^3 / \Gamma$.
The horizon radius $r_+$ is found from
\be 2\mu=r_+^4\left( 1-\frac{1}{r_+^2}\right)~. \ee The Hawking
temperature is \be\label{eqTH} T=\frac{2r_+^2-1}{2\pi r_+}~, \ee
while the mass and entropy of the black hole are given
respectively by
 \be\label{BH}
M=\frac{3V}{16\pi G}r_+^2(r_+^2-1) ~,~~~ S=\frac{V}{4G}r_+^3 \ee
where $V$ is the volume of the hyperbolic space $\Sigma_3$. Note that in the
horizon radius range $1/2\leq r^{2}_{+}\leq 1$ the mass of the black hole is 
negative~\cite{Mann:1997jb}. The lower bound corresponds to its maximum negative
value where the temperature is zero. The upper bound corresponds to zero
mass,  where as shown in~\cite{Koutsoumbas:2006xj} there is a phase transition of
the TBH to AdS space, while above that value the mass
takes positive values.

The energy of the dual CFT is \cite{Emparan:1999gf}
\be\label{ECFT} E_{CFT} = \frac{3V}{16\pi G } \left( r_+^2 -  \frac{1}{2} \right)^2 \ee
which is shifted with respect to the black hole energy by a positive amount 
 (Casimir energy due to counterterms one needs to add to the action to cancel infinities).
Notice that the minimum energy ($E_{CFT} = 0$) is at $T=0$, therefore the energy of the CFT is never negative, unlike its dual black hole.

For the study of perturbations, we need  the behaviour of harmonic
functions on $\Sigma_3$. In general, they obey
\be \left(\nabla^2 + k^2\right){\mathbb T}=0~.\label{angular}\ee
Without identifications (i.e., in $\mathbb{H}^3$), the spectrum is
continuous. We obtain \cite{TBH_perturbations}
%
\be\label{eqkxi} k^2=\xi^2+1+\delta \ee
where $\xi$ is arbitrary and $\delta =0,1,2$ for scalar, vector
and tensor perturbations, respectively. When a compactification
scheme is chosen, the spectrum becomes discrete. Depending on the
choice of $\Gamma$, the discretized eigenvalues $\xi^2$ may be made
as small as desired, i.e., zero is an accumulation point of the
spectrum of $\xi$ \cite{HyperB}. We also obtain negative values of $\xi^2$.
As $\xi^2$ approaches its minimum value, the complexity of
the set of isometries $\Gamma$ increases and the volume $V$ of the
hyperbolic space $\Sigma_3$ can be made arbitrarily large (hence also the mass and
entropy of the black hole).

Using the harmonics on $\Sigma_3$, we may write the wave equation
for gravitational perturbations in the general Schr\"odinger-like
form \cite{IK} \be\label{sch} -\frac{d^2\Phi}{dr_*^2} + V[r(r_*)]
\Phi = \omega^2 \Phi~, \ee in terms of the tortoise coordinate
$r_*$ defined by $ \frac{dr_*}{dr} = \frac{1}{f(r)} $ where $f(r)$
is defined in (\ref{metric}). The potential takes different forms
for different types of perturbation.


To calculate the Chern-Simons diffusion rate one needs to solve
the wave equation for a massless scalar field. The radial wave
equation is \be \frac{1}{r^{3}} ( r^{3} f(r) \Phi')' +
\frac{\omega^2}{f(r)} \Phi - \frac{k_\mathsf{S}^2}{r^2} \Phi = 0~.
\ee By defining $ \Phi = r^{-\frac{3}{2}} \Psi $ it can be cast
into the Schr\"odinger-like form (\ref{sch}) with the potential
given by \be\label{eqVV} V_\mathsf{S} (r) = f(r) \left\{
\frac{15}{4} + \frac{k_\mathsf{S}^2 - \frac{3}{4}}{r^2} +
\frac{9\mu}{2r^{4}} \right\}~. \ee
We may solve the wave equation
in terms of a Heun function and use the latter to determine the spectrum exactly albeit numerically \cite{Koutsoumbas:2006xj}.
However, such explicit expressions will not be needed for our purposes.

If the hyperbolic space $\Sigma_{3}$ is infinite, then
$k_\mathsf{S}^2 \ge 1$ (eq.~(\ref{eqkxi})). However, if
$\Sigma_{3}$ is finite, then it is easy to see that the minimum
eigenvalue is $ k_\mathsf{S}^2 = 0~. $ The corresponding
hyperspherical harmonic is a constant.
Above $k_\mathsf{S}^2=0$, the spectrum is
discrete.

For the AdS/CFT correspondence, we need the flux \be \mathcal{F} =
\frac{N^2}{16\pi^2}\ \sqrt{-g} g^{rr} \frac{\Phi^* \partial_r
\Phi}{|\Phi|^2} \Big|_{r\to\infty}~. \ee The imaginary part is
independent of $r$ (conserved flux). It is convenient to evaluate
it at the horizon where the wavefunction behaves as \be \Phi (r)
\approx \left( 1- \frac{r_+}{r} \right)^{-\frac{i\omega}{4\pi
T}}~. \ee We obtain \be \sqrt{-g} g^{rr} \Im (\Phi^* \partial_r
\Phi) = -\omega r_+^3 \ee therefore \be \Im\mathcal{F} = -
\frac{N^2 r_+^3\omega}{16\pi^2 |\Phi (\infty)|^2}~. \ee It is
related to the imaginary part of the retarded Green function, \be
\Im \tilde G^R (\omega, k_\mathsf{S}^2) = -2\Im \mathcal{F} \ee of
some scalar operator $\mathcal{O}$ ($G(x) = \langle \mathcal{O}
(x) \mathcal{O} (0) \rangle$, where $x\in \mathbb{R}\times
\Sigma_{3}$). We readily obtain \be \Im \tilde G^R(\omega,
k_\mathsf{S}^2) = \frac{N^2r_+^4\hat\omega}{8\pi^2
|\Phi(\infty)|^2}~. \ee For $\mathcal{O} = \frac{1}{4}
F_{\mu\nu}^a \tilde F^{a\mu\nu}$, we may define the Chern-Simons
diffusion rate
\be \Gamma = \left( \frac{g_{YM}^2}{8\pi^2} \right)^2 \int dt \int_{\Sigma_3} d^3\sigma \langle \mathcal{O} (x) \mathcal{O} (0) \rangle \ee
$\Gamma$ determines the rate of anomalous
baryon number violation at high temperatures in the Standard
Model. Expanding $G(x)$ in hyperspherical harmonics, the integral
over $\Sigma_{3}$ projects onto the lowest harmonic ($k_\mathsf{S}^2 =
0$). The integral over time then yields the Fourier transform at
$\omega = 0$. Using \be \tilde G(0,k_\mathsf{S}^2) = -
\lim_{\omega\to 0} \frac{2T}{\omega} \Im \tilde G^R
(\omega,k_\mathsf{S}^2) \ee we deduce \be \Gamma = \left(
\frac{g_{YM}^2}{8\pi^2} \right)^2 \tilde G (0,0) =
 \frac{(g_{YM}^2 N)^2}{256\pi^6}\ \frac{T r_+^3}{|\Phi(\infty)|^2} \Big|_{\omega = 0, k_\mathsf{S}^2 = 0} \ee
Evidently,
\be \Phi(r) \Big|_{\omega = 0, k_\mathsf{S}^2 = 0} = 1 \ee
at any temperature, therefore
\be \Gamma = \frac{(g_{YM}^2 N)^2}{256\pi^7}\ r_+^4 \left( 1 - \frac{1}{2r_+^2} \right)~. \ee
At high temperatures, $\Gamma \sim T^4$ whereas as $T\to 0$, $\Gamma \sim T \to 0$, i.e., anomalous baryon number violation is suppressed at low temperatures.


To calculate the shear viscosity, we need to discuss vector
gravitational perturbations. The lowest eigenvalue of the angular
equation (\ref{angular}) for a vector harmonic  $\mathbb{V}_i$
($\nabla_i \mathbb{V}^i = 0$) on a finite hyperbolic space
$\Sigma_{3}$ is found by observing that
\[ \nabla^j (\partial_i \mathbb{V}_j - \partial_j \mathbb{V}_i) = (k_\mathsf{V}^2 + 2)\mathbb{V}_i \]
where we used $R_{ij} = -2 \gamma_{ij}$ ($\gamma_{ij}$ being the
metric on $\Sigma_{3}$). Therefore, we have a constant vector
harmonic if $k_\mathsf{V}^2 + 2 =0$. The minimum eigenvalue is
$k_\mathsf{V}^2 = -2$. Above it, we have a discrete spectrum of
eigenvalues \be k_\mathsf{V}^2 = -2 + \Delta \ \ , \ \ \ \ \Delta
\ge 0~. \ee The radial wave equation is of the form (\ref{sch})
with potential \be V_\mathsf{V}(r) = f(r) \left\{ \frac{3}{4} +
\frac{k_\mathsf{V}^2 -\frac{7}{4}}{r^2} - \frac{27\mu}{2r^4}
\right\}~. \ee We may solve the radial equation and obtain a
solution in terms of a Heun function. Since we are interested in
the hydrodynamic behaviour, we shall solve the radial equation
only for small $\omega$ and $\Delta$ using perturbation theory.

More precisely, the hydrodynamic approximation is valid provided
\be\label{eqhc} \omega \ , \ \sqrt{\Delta} \ll r_+ \ee (recall
that we are working in units in which the AdS radius $l=1$). At
high temperatures, this constraint is equivalent to $\omega \ ,
\sqrt{\Delta} \ll T$. Also, the area of the horizon ($A_+\sim
r_+^2$)  is large and the constraint (\ref{eqhc}) is satisfied for
eigenvalues $\Delta \sim \mathcal{O} (1)$ because then $\Delta \ll
A_+$. This is similar to the case of a sphere. In both cases, the
hydrodynamic limit is valid at high temperature (large black hole)
\cite{Friess:2006kw,Alsup:2008fr}.

At low temperatures, in the case of a spherical horizon, its area
becomes small.  Even with $A_+\sim \mathcal{O} (1)$, it is no
longer possible to satisfy the constraint (\ref{eqhc}) because the
low-lying eigenvalues $\Delta \sim \mathcal{O} (1)$ regardless of
the size of the horizon. Thus, for a small spherical black hole
the hydrodynamic approximation is invalid.

For a hyperbolic horizon at low temperature, we have $r_+
\sim \mathcal{O} (1)$ ($r_+ \ge 1/\sqrt{2}$ at all
temperatures), so the hydrodynamic constraint (\ref{eqhc}) is not
always satisfied, as in the case of a spherical horizon. However, unlike in the case of a sphere, a hyperbolic space $\Sigma_3$ of high genus can have a large volume $V\gg 1$. The low lying eigenvalues are
\be \sqrt\Delta \sim \frac{1}{V^{1/3}} \ee
and therefore can be small ($\sqrt\Delta \lesssim \mathcal{O} (1)$) if $V$ is large.
Thus, for topological black holes of high genus hyperbolic horizons
the hydrodynamic approximation is valid even in the low
temperature (small horizon radius) limit owing to the complexity of the horizon surface.

To solve the radial wave equation, it is convenient to introduce the coordinate
\be u = \left(
\frac{r_+}{r} \right)^2~. \ee
In terms of the wavefunction $F(u)$
defined by \be \Psi(u) = (1-u)^{- \frac{i\omega}{4\pi T}} \ F(u)
\ee we have \be \mathcal{A} F'' + \mathcal{B} F' + \mathcal{C} F =
0~, \ee where
\bea \mathcal{A} &=& u\hat f~, \nonumber\\
\mathcal{B} &=& u\hat f' + \frac{3}{2} \hat f + \frac{i\omega}{4\pi T} \frac{u\hat f}{1-u}~, \nonumber\\
\mathcal{C} &=& -\frac{\hat V}{4u^2\hat f} + \frac{i\omega}{4\pi T} \frac{u\hat f' + \frac{3}{2} \hat f}{1-u} + \frac{i\omega}{4\pi T}\, \frac{u\hat f}{(1-u)^2} \nonumber\\
& & + \ \mathcal{O} (\omega^2/T^2)~. \eea
where prime denotes differentiation with respect to $u$ and
we have defined \bea\label{eq14} \hat f(u) \equiv
\frac{f(r)}{r_+^2} &=& \frac{1}{u} - \frac{1}{r_+^2} -
\frac{2\mu}{r_+^{4}} u~, \nonumber \\ \hat V_\mathsf{V}(u) \equiv
\frac{V_{\mathsf{V}}(r)}{r_+^2}&=& \hat f(u) \left\{ \frac{3}{4} +
\frac{\Delta -\frac{15}{4}}{r_+^2} u - \frac{27\mu}{2r_+^4}\, u^2
\right\}~.\nonumber\\ \eea
We obtain the zeroth
order equation by setting $\omega = 0, \Delta = 0$. The acceptable
solution is
\be F_0 = u^{3/4} \ee
independent of the temperature.

\begin{figure}[!b]
\centering
\includegraphics[angle=-90,scale=0.3]{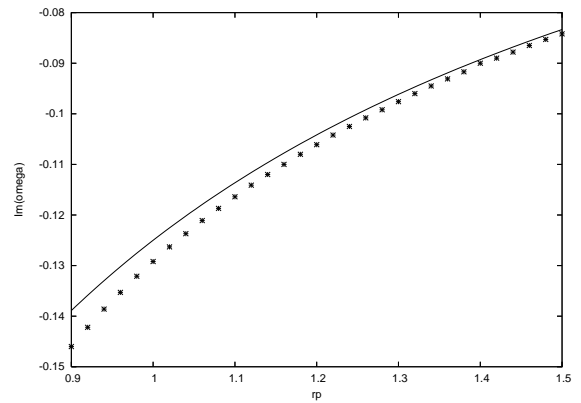}
\caption{ The imaginary part of the lowest, purely dissipative,
mode versus $r_{+}$. The continuous line and the points represent
the perturbative and numerical results, respectively.}
\label{05}
\end{figure}

Expanding the wavefunction,
\be F = F_0 + F_1 + \dots~, \ee
at first order the wave equation reads
\be \mathcal{H}_0 F_1 = - \mathcal{H}_1 F_0~, \ee
where
\be
\mathcal{H}_1 F_0 = \frac{i\omega}{4\pi T} \left\{ \frac{2}{u} + 3
\left( 1 - \frac{1}{r_+^2} \right) \right\} F_0 -
\frac{\Delta}{4r_+^2 u} F_0~. \ee
The solution may be written as
\be F_1 = F_0 \int \frac{\mathcal{W}}{F_0^2} \int \frac{F_0\mathcal{H}_1 F_0}{\mathcal{A}\mathcal{W}} ~, \ee
where
$\mathcal{W} = 1/(u^{3/2}\hat f)$ is the Wronskian.
The limits of the inner integral may be adjusted at will because this amounts to adding an arbitrary amount of the unacceptable zeroth-order wavefunction.
To ensure regularity at the horizon, we should choose one of the limits of integration at $u=1$. Then by demanding that the singularity vanish at the boundary ($u=0$), we arrive at the
first-order constraint \be \int_0^1 du \frac{F_0
\mathcal{H}_1 F_0}{\mathcal{A}\mathcal{W}} = 0 \ee 
After some straightforward algebra, this leads to the dispersion
relation \be\label{eqdisp} \omega = -i \frac{\Delta}{4r_+} \ee in
agreement with ref.~\cite{Alsup:2008fr} at high temperatures and
matching numerical results at all temperatures (Fig.~\ref{05}).
From (\ref{eqdisp}) we read off the diffusion coefficient
\be\label{eqD1} D = \frac{1}{4r_+} \ee which is related to the
viscosity coefficient via \be\label{eqD} D = \frac{\eta}{\epsilon
+ p}~. \ee This is known to be valid in flat space. It is also
valid in our case, as can be seen by writing the hydrodynamic
equations $ \nabla_\mu T^{\mu\nu} = 0 $ for a static fluid of
constant pressure perturbed by a small velocity field $ u^i =
e^{-i\omega t} \mathbb{V}^i~. $ The conservation law of the
hydrodynamic equations   yields~\cite{Alsup:2008fr} \be -4i\omega
p + \eta (k_\mathsf{V}^2 + 2) = 0~. \ee Eq.~(\ref{eqD}) then
follows if we use (\ref{eqdisp}), (\ref{eqD1}) together with
$\epsilon = 3p$ which is valid for a confomal fluid.

From the expression for the energy (\ref{ECFT}), we obtain the
energy density $\epsilon = E_{CFT}/V$ and the shear viscosity
coefficient
\be \eta = \frac{4}{3} \epsilon D = \frac{1}{16\pi G r_+} \left( r_+^2 - \frac{1}{2} \right)^2 \ . \ee
Dividing by the
entropy density ($s = S/V$, where the entropy is given by
(\ref{BH})), we obtain \be \frac{\eta}{s} = \frac{1}{4\pi} \left(
1 - \frac{1}{2r_+^2} \right)^2~. \ee At high temperature (large
$r_+$), $\frac{\eta}{s} \approx \frac{1}{4\pi}$. As $T\to 0$,
$\frac{\eta}{s} \sim T^2 \to 0$. At all temperatures, the ratio is
below $\frac{1}{{4\pi}}$.

In flat spacetime, one also obtains the viscosity coefficient from the Kubo formula \cite{Hydrodynamics}, which agrees with the result obtained via the diffusion coefficient. The former is derived by considering tensor perturbations.
This is not possible in our case, because tensor perturbations of the static fluid in hyperbolic space do not exist due to their being traceless and divergenceless \cite{Alsup:2008fr}.

\textit{In conclusion} Using the AdS/CFT correspondence we have
calculated the anomalous baryon number violation rate and the
ratio of shear viscosity to entropy density in thermal field
theories having gravity duals with hyperbolic horizons. We found
the explicit temperature dependence of the anomalous baryon number
violation and we showed that it is suppressed at low temperatures.
For high genus hyperbolic spaces the hydrodynamic approximation is
valid at low temperatures, and the ratio of shear viscosity to
entropy density is found to be below $1/(4\pi)$ at all
temperatures. It can be made arbitrarily small in the low
temperature limit.

\textit{Acknowledgments} We thank J.~Nat\'ario for discussions.
Work supported by the NTUA research program PEVE07. G.~S.~was
supported in part by the US Department of Energy under grant
DE-FG05-91ER40627.


\begin{thebibliography}{99}


\bibitem{Hydrodynamics}
  G.~Policastro, D.~T.~Son and A.~O.~Starinets,
  Phys.\ Rev.\ Lett.\  {\bf 87}, 081601 (2001);
  G.~Policastro, D.~T.~Son and A.~O.~Starinets,
  JHEP {\bf 0209}, 043 (2002);
  P.~Kovtun, D.~T.~Son and A.~O.~Starinets,
  JHEP {\bf 0310}, 064 (2003).



\bibitem{Kovtun:2004de}
  P.~Kovtun, D.~T.~Son and A.~O.~Starinets,
  Phys.\ Rev.\ Lett.\  {\bf 94}, 111601 (2005).

\bibitem{chemical}
  J.~Mas,
  JHEP {\bf 0603}, 016 (2006);
  D.~T.~Son and A.~O.~Starinets,
  JHEP {\bf 0603}, 052 (2006);
  O.~Saremi,
  JHEP {\bf 0610}, 083 (2006);
  K.~Maeda, M.~Natsuume and T.~Okamura,
  Phys.\ Rev.\  D {\bf 73}, 066013 (2006).

\bibitem{corrections}
  A.~Buchel, J.~T.~Liu and A.~O.~Starinets,
  Nucl.\ Phys.\  B {\bf 707}, 56 (2005);
  P.~Benincasa and A.~Buchel,
  JHEP {\bf 0601}, 103 (2006).
  R.~C.~Myers, M.~F.~Paulos and A.~Sinha,
  Phys.\ Rev.\ D {\bf 79}, 041901(R) (2009).

\bibitem{Cai:2008in}
  R.~G.~Cai and Y.~W.~Sun,
  JHEP {\bf 0809}, 115 (2008).

\bibitem{GB_correction}
  M.~Brigante, H.~Liu, R.~C.~Myers, S.~Shenker and S.~Yaida,
  Phys.\ Rev.\  D {\bf 77}, 126006 (2008);
  M.~Brigante, H.~Liu, R.~C.~Myers, S.~Shenker and S.~Yaida,
  Phys.\ Rev.\ Lett.\  {\bf 100}, 191601 (2008);
  Y.~Kats and P.~Petrov,
  JHEP {\bf 0901}, 044 (2009).

\bibitem{Horowitz:1991cd}
  G.~T.~Horowitz and A.~Strominger,
  Nucl.\ Phys.\  B {\bf 360}, 197 (1991);
  G.~W.~Gibbons and P.~K.~Townsend,
  Phys.\ Rev.\ Lett.\  {\bf 71}, 3754 (1993).



\bibitem{Kehagias:2000dga}
  A.~Kehagias and J.~G.~Russo,
  JHEP {\bf 0007}, 027 (2000).

\bibitem{Emparan:1999gf}
  R.~Emparan,
  JHEP {\bf 9906}, 036 (1999).

\bibitem{Son:2002sd}
  D.~T.~Son and A.~O.~Starinets,
  JHEP {\bf 0209}, 042 (2002);
  G.~Policastro, D.~T.~Son and A.~O.~Starinets,
  JHEP {\bf 0209}, 043 (2002).

\bibitem{Friess:2006kw}
  J.~J.~Friess, S.~S.~Gubser, G.~Michalogiorgakis and S.~S.~Pufu,
  JHEP {\bf 0704}, 080 (2007);
  G.~Michalogiorgakis and S.~S.~Pufu,
  JHEP {\bf 0702}, 023 (2007).

\bibitem{Alsup:2008fr}
  J.~Alsup and G.~Siopsis,
  Phys.\ Rev.\ D {\bf 78}, 086001 (2008).

\bibitem{topological}
  R.~B.~Mann,
  Class.\ Quant.\ Grav.\  {\bf 14}, L109 (1997);
  R.~B.~Mann,
  Nucl.\ Phys.\ B {\bf 516}, 357 (1998);
  L.~Vanzo,
  Phys.\ Rev.\ D {\bf 56}, 6475 (1997);
  D.~R.~Brill, J.~Louko and P.~Peldan,
  Phys.\ Rev.\ D {\bf 56}, 3600 (1997);
  D.~Birmingham,
  Class.\ Quant.\ Grav.\  {\bf 16}, 1197 (1999).

\bibitem{TBH_perturbations}
  J.~S.~F.~Chan and R.~B.~Mann,
  Phys.\ Rev.\  D {\bf 59}, 064025 (1999);
  B.~Wang, E.~Abdalla and R.~B.~Mann,
  Phys.\ Rev.\  D {\bf 65}, 084006 (2002);
  R.~Aros, C.~Martinez, R.~Troncoso and J.~Zanelli,
  Phys.\ Rev.\  D {\bf 67}, 044014 (2003);
  D.~Birmingham and S.~Mokhtari,
  Phys.\ Rev.\  D {\bf 74}, 084026 (2006).




\bibitem{Koutsoumbas:2006xj}
  G.~Koutsoumbas, S.~Musiri, E.~Papantonopoulos and G.~Siopsis,
  JHEP {\bf 0610}, 006 (2006);
  G.~Koutsoumbas, E.~Papantonopoulos and G.~Siopsis,
  JHEP {\bf 0805}, 107 (2008);
  G.~Koutsoumbas, E.~Papantonopoulos and G.~Siopsis,
  arXiv:0806.1452 [hep-th].

\bibitem{Mann:1997jb}
  R.~B.~Mann,
  Class.\ Quant.\ Grav.\  {\bf 14}, 2927 (1997)
  [arXiv:gr-qc/9705007].

\bibitem{HyperB}
 N.~L.~Balazs and A.~Voros,
  Phys. Reports {\bf 143(3)}, 109 (1986).
  N.~J..~Cornish and D.~N.~Spergel,
  arXiv:math/9906017;
    K.~Inoue,
  Class.\ Quant.\ Grav.\  {\bf 16}, 3071 (1999).

 \bibitem{IK}
A.~Ishibashi and H.~Kodama,
Prog.~Theor.~Phys.~{\bf 110}, 701 (2003).





\end{thebibliography}
\end{document}